\documentclass{iopart}
\usepackage{amssymb}
\usepackage{graphicx,subfigure}
\begin{document}

\newcommand{\bra}[1]{\left\langle #1 \right |}
\newcommand{\ket}[1]{\left | #1 \right\rangle}
\newcommand{\braket}[2]{\left\langle #1 | #2 \right\rangle}
\newcommand{\Matrix}[2]{\left( \begin{array}{#1} #2 \end{array}
  \right)}
\newcommand{\bigchoose}[2]{\mbox {${\displaystyle{ #1 \choose #2}}$}}
\newcommand{\expect}[1]{\left\langle #1 \right\rangle}
\def\id{{\rm 1\kern-.22em l}}
\def\trace{{\rm tr}\;}

\title{Robustness of adiabatic passage through a quantum phase transition}
\author{Andrea Fubini$^{1,2,3}$, Giuseppe Falci$^3$, and Andreas Osterloh$^4$}
\address{$^1$CNISM, Unit\`a di Firenze, Via G. Sansone 1,
    I-50019 Sesto Fiorentino (FI), Italy}
\address{$^2$Dipartimento di Fisica, Universit\`a degli Studi di Firenze, 
	Via G. Sansone 1, I-50019 Sesto Fiorentino (FI), Italy}
\address{$^3$MATIS-INFM $\&$ Dipartimento di Metodologie Fisiche e
    Chimiche (DMFCI), Universit\`a di Catania, viale A. Doria 6, I-95125 Catania,
    Italy}
\address{$^4$Institut f\"ur Theoretische Physik,
Leibniz Universit\"at Hannover, Appelstrasse 2, D-30167 Hannover, Germany}

\begin{abstract}
We analyze the crossing of a quantum critical point based on
exact results for the transverse XY model. In dependence of the change rate
of the driving field, the evolution of the ground state 
is studied while the transverse magnetic field is tuned through the 
critical point with a linear ramping. The excitation probability is obtained
exactly and is compared to previous studies and to the Landau-Zener
formula, a long time solution for non-adiabatic transitions in two-level systems.
The exact time dependence of the excitations density in the system allows to identify the adiabatic and diabatic regions during the sweep and to study the mesoscopic fluctuations of the excitations.
The effect of white noise is investigated, where the critical point transmutes
into a non-hermitian ``degenerate region''.
Besides an overall increase of the excitations during and at the end of the sweep, the most destructive effect of the noise is the decay of the state purity that is enhanced by the passage through the degenerate region. 
\end{abstract}

\pacs{73.43.Nq 05.40.-a 75.10.Jm}

\maketitle

\section{Introduction}
Adiabatic passages are a fascinating and important tool in modern physics,
since they allow to manipulate quantum states in a controlled manner
via tunable parameters of a physical model system. Their potential
impact ranges from setting current standards by means of adiabatically
pumping charge through a quantum dot\cite{AdiabatPumps1,AdiabatPumps2,AdiabatPumps3,AdiabatPumps4}
to the access to ground states
of e.g. non-integrable Hamiltonians via adiabatically connecting to it 
from an integrable Hamiltonian\cite{AdiabatQComp}.
Since modern methods in quantum optics
allow fabrication of a wide class of lattice Hamiltonians with
systems of atoms loaded into a suitably engineered optical trap,
these techniques have become experimentally accessible nowadays.
We here will discuss the latter application of a possibly adiabatic
transit of a ground state with an explicit time dependent Hamiltonian
and ``slowly'' varying parameters, which in the recent literature
has also been termed {\em adiabatic quantum computation} (AQC)\cite{AdiabatQComp}.
The efficiency of this procedure is compromised by size effects and by
the presence of noise, the arch-enemy of all quantum information tasks.

The closing of the gap in the spectrum of excitations with growing system 
size gives a limit to the rate the parameters may be changed in order
to have a sufficiently high fidelity with the ground state of the
final Hamiltonian. In other words, the gap rules the time of computation.
An intriguing question is then, what happened if
a quantum critical point\cite{SACHDEV} is crossed during the passage.
In the dynamic scenario presented here, the vanishing gap makes the notion of
slow change rates obsolete, and strictly speaking no adiabatic passage
exists across a quantum critical point.
In the laboratory we have to deal with finite systems
and as a consequence the gap, though scaling down to zero somehow with
the system size, will nevertheless be finite. Then, the question arises
how the fidelity of the ground state and/or the time of computation
scales with the system size. This question is relevant for deciding
whether adiabatic quantum computation may be successful or not.
It has been addressed for
the one-dimensional transverse Ising and the anisotropic Heisenberg-chain\cite{MurgCirac} in
numerically solving the Heisenberg equation;
the excitation probability has been calculated perturbatively taking into account a finite number of multiparticle excitations.
The authors concluded that the adiabatic time scale for the Ising model in a transverse field scales polynomially with the system size. 
Besides a complete treatment of many-particle excitations 
this work did not take into consideration the crucial role fluctuations of the
average excitation number play in particular for small system size.
The previous question is also related to the Kibble-Zurek theory, where critical
scaling laws have been employed to obtain predictions for
the density of defects after the crossing of a
real phase transition\cite{Kibble,Zurek1,Zurek2}. Its
predictions have been tested for the one-dimensional transverse
Ising model in Ref.~\cite{ZurekDornerZoller},
where the kink density has been calculated numerically and good qualitative
agreement with the Kibble-Zurek scenario has been reported up to little
quantitative discrepancy. 
By means of simple scaling arguments the density of created excitations as function of the sweeping rate has been evaluated for models belonging to different universality classes\cite{Polkovnikov05}.
Using the exact solution for the
transverse Ising model~\cite{McCoy1,LiebSchulzMattis,Pfeuty,McCoy2}, 
the kink density at vanishing magnetic field has been calculated 
in Ref.~\cite{Dziarmaga05}, and numerical discrepancies with 
Ref.~\cite{ZurekDornerZoller} have been noted, 
where the time dependent Bogoliubov approach
has been applied and excitation probabilities in the long time limit 
are obtained from a mapping to Landau-Zener tunneling. 

How the presence of the noise affects this scenario is a rather unexplored,
though very important, topic. In experimental setups one has to struggle 
with the influence of environmental degrees of freedom. This transforms 
the avoided criticality of the finite system
into a region of non-hermitian degeneracy\cite{berry} 
extending over a finite interval in control parameter space.
This means that the system passes through a gapless phase within 
a finite time window. Moreover, besides the inverse of the spectral 
gap that sets a lower bound on the sweeping period, 
the noise introduces a second time scale: 
the dephasing time, that on the contrary sets an upper bound.
These competing time scales give rise to non-trivial effects,
which are largely ignored in the recent literature on 
AQC~\cite{Childs,Aberg,Roland} 
notwithstanding their importance for a realistic analysis of AQC protocols.

A prominent and simple class of models with quantum critical point are
the one dimensional XY models in transverse magnetic field\cite{SACHDEV,McCoy1}.
In this work, we trace the exact evolution of the model during a 
sweep in order to study the robustness of the adiabatic passage against 
size effects with and without the presence of noise in the 
driving field. The paper is organized as follows: In section \ref{model} 
we present the model Hamiltonian together with a sketch of its 
exact solution of the dynamics and we briefly review the Kibble-Zurek 
scenario for a quantum phase transition in section \ref{KZ}. 
In section \ref{fluct}, 
the results for the excitation probability and its mean square 
fluctuations during the sweep are presented and discussed. 
In section \ref{noise}, we analyze the case of a noisy driving field. 
Conclusions out of our findings are drawn in section \ref{concl}.

\section{The model and its exact solution}\label{model}

The model Hamiltonian under consideration throughout this work
is the one-dimensional spin-1/2 XY model in transverse magnetic field,
an archetype model exhibiting a quantum phase transition\cite{SACHDEV}
for which an exact solution exists~\cite{McCoy1,LiebSchulzMattis,Pfeuty,McCoy2}
\begin{equation}
\label{Hamiltonian}
H=-\left[\sum_{j=1}^{L} (1+\gamma) S^x_j S^x_{j+1}
        + (1-\gamma) S^y_j S^y_{j+1}\right]
    - h(t) \sum_{j=1}^{L} S^z_j\;.
\end{equation}
>From here on, for sake of simplicity the unit of energy, $\hbar$ and the lattice spacing are set to one. In the above equation $\gamma$ is the in-plane anisotropy coefficient, the magnetic
field $h(t)$ is the time-dependent tunable parameter 
and the boundary conditions are periodic. 
The model
undergoes a quantum phase transition at $h_{\rm c}=1$ separating
a paramagnetic ($h>h_{\rm c}$) from a broken symmetry phase, where
$\langle S^x \rangle \neq 0$.  Details about the exact solution and
the derivation of the exact time evolution 
are found in \cite{McCoy1}, and the main steps are outlined in the Appendix.
The canonical transformations (\ref{hard-core-bosons}),
(\ref{JW}) and (\ref{FT}) completely decouple the Hamiltonian into a
direct sum $H=\oplus_{|k|} H_k$ of 4-dimensional Hamiltonians $H_k$
acting non-trivially only in the Hilbert space
$\{\ket{0}_{k,-k},\ket{k,-k};\ket{k},\ket{-k}\}$, where, due to the
parity symmetry, it further decouples into $H_k=H_k^{odd}\oplus
H_k^{even}$. Whereas in the odd-occupation Hilbert space the
Hamiltonian is already diagonal, $H_k^{odd}=-\cos{\phi_k}\id_2 $, in
the even-occupation Hilbert space and basis $\{\ket{0},\ket{k,-k}\}$
one obtains
\begin{eqnarray}
\label{H^e_pauli} H_k^{even}=-\cos{\phi_k} \id_2 + a_k \sigma^z +
b_k \sigma^y\,,
\end{eqnarray}
where $\phi_k=2\pi k/(L-1)$, $a_k=\cos \phi_k + h(t)$, 
$b_k=\gamma \sin \phi_k$, and $\sigma^\alpha$ are the Pauli matrices. 
>From now on we will focus on the nontrivial even part of
the Hamiltonian and for sake of simplicity the superscript ``even'' will be
dropped. The eigenvalues are
\begin{equation}
\label{Ek}
E_k=\pm\Lambda_k=\pm
\sqrt{a_k^2+ b_k^2}
\end{equation}
with $a_k$ and $b_k$ as defined above, and the ground state is
\begin{equation}
\label{e.GS} \ket{GS_k}=\bigchoose{-i \beta_k}{\alpha_k}=:
    (-i\beta_k + \alpha_k f^\dagger_k f^\dagger_{-k}) \ket{0}_{k,-k}\; ,
\end{equation}
where
\begin{equation*}
\alpha_k = \frac{\Lambda_k+a_k}{\sqrt{2\Lambda_k(\Lambda_k+a_k)}}\,,
~~~~~~~\beta_k = \frac{b_k}{\sqrt{2\Lambda_k(\Lambda_k+a_k)}}\,,
\end{equation*}
and the fermionic operators $f_l^\dagger$ are defined in
Eq.(\ref{FT}). The ground state of the whole chain, which later on will be
the initial state, is the tensor product
$$\ket{GS}=\prod^\otimes_{|k|}\ket{GS_k}\; ,\quad
\rho_{_{t=0}}=\prod^\otimes_{|k|}\ket{GS_k}\bra{GS_k}\; ,$$
since we will start in the phase where the parity symmetry is
not broken. As a consequence, $\expect{S^x}=0$ during the time evolution.

In terms of single mode operators the Hamiltonian can be
written as
\begin{equation}
H=\sum_k \Lambda_k (\eta_k^\dagger \eta_k - \frac12) ~,
\label{e.Heta}
\end{equation}
where the relation between the fermionic operators $\eta_k$ and $f_k$
reads
\begin{equation*}
\eta_k = \beta_k f_k + i \alpha_k f^\dagger_{-k} ~.
\end{equation*}

It is worth noticing that the diagonalization is not giving the time
evolution, since $[\partial_t H, H]\neq 0$.
The time evolution operator is given by
\begin{equation}\label{U}
U(t):={\rm T}\exp\{-i\int_0^t{\rm d\tau}H(\tau)\}
\end{equation}
with T the time ordering operator, and we have to deal with the
Schr\"odinger equation
\begin{equation}
\label{e.HE} i\partial_t U=HU\; ;\quad U=\prod^\otimes_{|k|}
U^{(k)}\, ,
\end{equation}
for the time evolution operator $U$ instead.
An exact solution does exist for several time
dependences~\cite{McCoy1,zener,VOROS}. In the simplest case
of linear sweep $h(t)=h_0+(h_1-h_0)t/T$ the long time physics 
exhibits the Landau-Zener tunneling scenario~\cite{zener,landau}; 
for details of the exact time evolution we refer to the Appendix.

The fidelity of an adiabatic passage process is the overlap of the
wavefunction at time $t$ and the ground state $\ket{GS(t)}$ of $H(t)$
at the same time. The latter would be the wavefunction for
adiabatically driven magnetic field. The overlap integral
$\bra{GS(t)}U(t)\ket{GS(0)}$ is thus a rough measure for the efficiency
of the adiabatic passage. Another way to check the adiabaticity of a
sweeping process is to evaluate the average number of excitations in
the system after the sweep, i.e. the expectation value of $\sum_k
\eta_k^\dagger \eta_k$ on the state $U(t)\ket{GS(0)}$. The
passage is fully adiabatic if no excitations are produced at the end
of the drive.

\section{Kibble-Zurek scenario}\label{KZ}

The density of excitations crossing a phase transition can be
foreseen according to the Kibble-Zurek
scenario~\cite{Kibble,Zurek1,Zurek2}. Before entering the bulk of our work, we shall give a short summary
of results from the recent paper by Zurek, Dorner and Zoller (ZDZ)~\cite{ZurekDornerZoller}, who extended the Kibble-Zurek scenario to quantum critical phenomena.
Let us assume a linear sweep of
the field in a period $T$ between $h_0$ and $h_1$: $$ h(t)=h_0-v_h
t~,~~~~~~~~~v_h\equiv\frac{h_0-h_1}{T}~.  $$ 
We define the distance from
the critical point, $\epsilon(t)=h(t)-h_{\rm c}$. From the
dispersion relation, Eq.~(\ref{Ek}), it follows that
in the thermodynamic limit the gap in the excitation spectrum is
$$
\Delta = |h - h_{\rm c}|~,
$$ while in a finite chain with $L$ spins at $h=h_{\rm c}$ the gap 
takes the finite minimum value
$$
\Delta_L = \frac{2\pi \gamma}{L} +O\left(L^{-2}\right)~.
$$
$\Delta$ (or $\Delta_L$) sets the energy scale and 
the {\em relaxation time} is
$$
\tau = \frac{\tau_0}{\Delta}
=\frac{\tau_0}{|\epsilon(t)|} ~,
$$
where $\tau_0$ is an appropriate coefficient of proportionality.
The divergence of $\tau$ is the hallmark of the critical slowing down
in the usual phase transition. One can also define a {\em healing
length}: $\xi=c\tau$, where $c=\displaystyle{\lim_{k\to\pi}}
\frac{\partial \Lambda_k}{\partial k}\bigg|_{h=1}\!\!=\gamma$ is the spin wave velocity,
and a density of excitations $\nu = 1/\xi$.

According to ZDZ, when approaching the critical point,
the dynamics of the system ideally should change from adiabatic to 
frozen behavior at time $\hat t$. Since $\epsilon(t)$ changes on a
time scale $\epsilon(t)/\dot{\epsilon}(t)$ the crossover between
an adiabatic and a sudden region should occur when~\footnote{Here
the time set-off is chosen such that the critical point is reached at $t=0$.} 
$$
\tau(\hat t) = \frac{\epsilon(\hat t)}{\dot{\epsilon}(\hat t)}=\hat t
~~~~~~\Rightarrow ~~~~~~\hat{t}^2 = \tau_0/v_h~.
$$
When crossing the critical point at velocity $v_h$,
the healing length at $\hat{t}$ is $\hat\xi=c\hat t$ and 
the excitations density after the sweep would be $\nu \propto \sqrt{v_h}$. 
Studying the time dependence of different quantities in the system we will
directly probe this scenario.

\section{Effect of mesoscopic fluctuations on the ZDZ scaling law}\label{fluct}

\begin{figure}
\centering
\includegraphics[width=6cm]{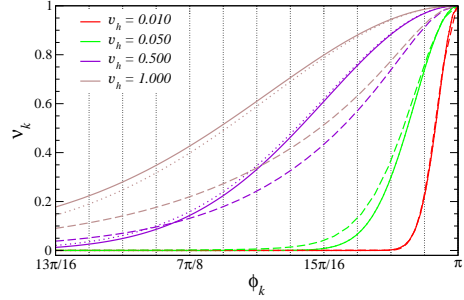}
\caption{\label{f.eta_k}
Single mode excitation density vs. $\phi_k$ for $\gamma=1$ at different quench rates (from
bottom to top: $v_h = 0.01,0.05,0.50,1.00$) and different final fields
$h_{\rm f}=0$ (solid lines) and $h_{\rm f}=0.9$ (dashed). Dotted lines
represent the predictions from the Landau-Zener formula. Vertical lines 
point out the values of $\{\phi_k\}$ for $L=129$.}
\end{figure}
After a sweep from $h(0)$ to $h(t)$ the density of excitations at
time $t$ in the system is given by the expectation value
\begin{eqnarray}
\nu &=\frac1L \langle GS | U^\dagger(t)  \sum_k \eta_k^\dagger \eta_k \,U(t)
| GS \rangle \nonumber\\
& =\frac2L\sum_{k>0}\big\{\alpha_k^2 - (\alpha_k^2 - \beta_k^2 )
|\alpha_k(t)|^2 - 2 \alpha_k\beta_k \, \Im[ \alpha_k^*(t) \beta(t) ]\big\}
\nonumber\\
&\equiv \frac{1}{L}\sum_{k\neq 0} \nu_k ~, \label{e.nu_k}
\end{eqnarray}
where $\eta^\dagger$ ($\eta$) is the single mode creation (annihilation) operator as defined by Eq.~(\ref{e.Heta}) and $U(t) | GS \rangle \equiv \prod^\otimes_{|k|}
[-i\beta_k(t)+\alpha_k(t)f^\dagger_k f^\dagger_{-k}]\ket{0}_{k,-k}$
~\footnote{ It is worth noticing that the $k=0$ mode can only be 
singly occupied or empty. Its occupation corresponded to the odd sector, 
and therefore the $k=0$ mode must be excluded.}.
At $\gamma=1$ and $h(T)=0$ we have $\sum_k \eta_k^\dagger
\eta_k = \frac12\sum_i(1-\sigma^x_i\sigma^x_{i+1})$, so that $\nu$
coincides with the kink density in the system at the 
end of the sweep~\cite{Dziarmaga05}.

Fig.\ref{f.eta_k} shows the excitation spectrum for
different velocities. For a finite chain of $L=2N+1$ sites the
lowest energy mode is $\phi_{N-1}=\pi(N-1)/N$, this means that, for
instance, at $v_h=0.01$ the sweep can be considered adiabatic up to
$L\lesssim 65$. Looking at Fig.\ref{f.eta_k} one can compare the
exact result and the Landau-Zener formula $P_k = \exp[-2\pi\gamma^2
\frac{\sin^2(\phi_k)}{v_h}]$ that reproduces surprisingly well the
exact solution for final field $h_{\rm f}=0$ even for rather high $v_h$. This
confirms that the physics of the process is ruled by the Landau-Zener
tunneling~\cite{Dziarmaga05}.  
One can also observe that for sufficiently low
sweeping rate the excitation density just after the transition point
($h=0.9$) is the same as found at $h=0$.
\begin{figure}
\centering
 \subfigure
{\includegraphics[width=6cm,height=55mm]{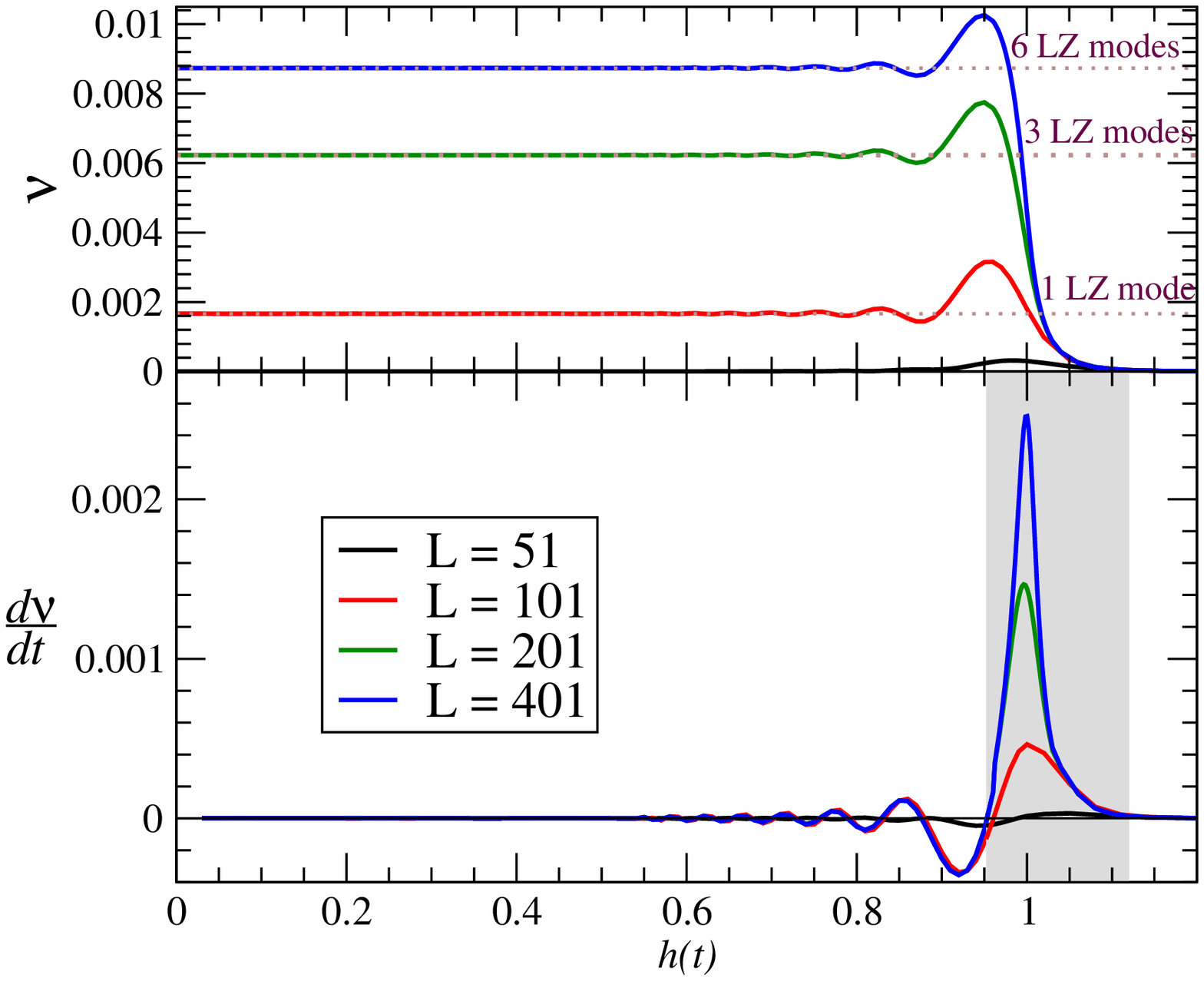}}
 \subfigure
{\includegraphics[width=6cm,height=55mm]{fig2right.eps}}
\caption{\label{f.region} Left upper panel: Density of excitations  vs.
$h(t)=h_0-v_h t$ for different chain lengths (from bottom to top:
$L=51,101,201,401$) at $v_h=0.01$ and $\gamma=1$. 
Dotted lines represent the prediction
from the Landau-Zener formula and a finite number of modes. Left
bottom panel: Time derivative of the excitation density. 
The shaded region
points out the ``sudden interval'' in the sweep.  Right panel:
$\frac{d\nu}{dt}$ vs. $h(t)$ for $\gamma=1$ and increasing sweeping
velocities $v_h$ (from bottom to top). The shaded areas point out the
sudden regions that broadens with increasing $v_h$.}
\end{figure}

An adiabatic and a sudden region can be distinguished:
when the system enters the sudden region the number of excitations rapidly
increases up to some maximum value. Afterwards, an oscillatory behavior
of the excitation density is observed.
Thus, a sensitive (though not unique) possibility to identify
the sudden region is the time derivative of the excitation density
$\frac{{\rm d}\nu}{{\rm d}t}=-v_h \frac{{\rm d}\nu}{{\rm d}h}$.
Using this method the sudden region can be depicted as the grey shaded
regions in Fig.\ref{f.region}. At fixed $v_h$, the
number of excitations clearly depends on the density of states (DOS)
between the ground state energy and the ``sweeping energy''
$\epsilon_v:=v_h$. 
In particular only few modes contribute
to the total excitation density and their number is related to
the minimum gap $\Delta_L=2\pi\gamma/L$ and to the DOS 
(upper left panel of Fig.~\ref{f.region}).
When the sweeping velocity increases the sudden region broadens 
(Fig.~\ref{f.region}, right panel). 
The Kibble-Zurek scenario predict the width of this region to be proportional to the
squareroot of the sweeping velocity: $\Delta\hat h \propto v_h \hat t =
\sqrt{\tau_0 v_h}$.
\begin{figure}
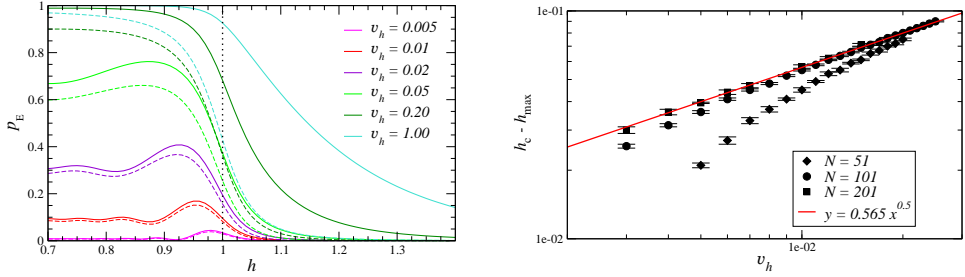

 \centering
 \subfigure
   {\includegraphics[width=6cm]{fig3left.eps}}
 \hspace{5mm}
 \subfigure
{\includegraphics[width=6cm]{fig3right.eps}}
 \caption{\label{f.scaling} Left panel: $p_{_{\rm E}}$ for
different quench rates $v_h$, for  $\gamma=1$ and $L=101$.
Full lines represent the total $p_{_{\rm E}}$, and the dashed
lines represent the probabilities to excite the first mode only. Right
panel: scaling of $\Delta\hat h$ vs $v_h$ as $N$ increases on a
log-log plot.}
\end{figure}
Already a much less detailed quantity reveals similar insight 
and admits comparison with reference~\cite{MurgCirac}; namely the total 
excitation probability
\begin{equation}
p_{_{\rm E}} = 1 - |\bra{GS(t)}U(t)\ket{GS(0)}|^2~.
\end{equation}
Like $\nu$, it exhibits a maximum just after the crossing of
critical point (see Fig.~\ref{f.scaling}). 
We define $\Delta\hat h \equiv h_{\rm c} - h_{\rm max}$,
where $h_{\rm max}$ is the value of the field where $p_{_{\rm E}}$
reaches its maximum. The value of $\Delta\hat h$ evaluated in this
way depends on $L$ due to the $L$-dependence of the gap. 
It is only the extrapolation of $\Delta\hat h$ to $L\to\infty$ that 
clearly displays the Kibble-Zurek square root behavior $\sqrt{\tau_0 v_h}$ 
as can be seen from the log-log plot in Fig.~\ref{f.scaling}.

\begin{figure}
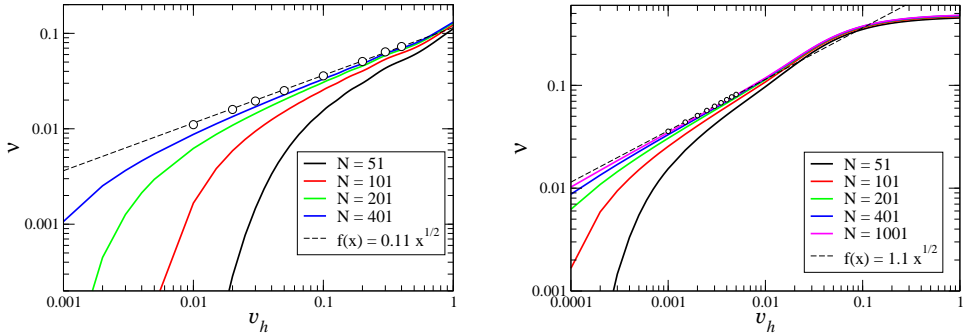

 \centering
 \subfigure
   {\includegraphics[width=6cm]{fig4left.eps}}
 \hspace{5mm}
 \subfigure
{\includegraphics[width=6cm]{fig4right.eps}}
 \caption{\label{f.nu_vs_vh} Left panel: Excitation density vs. $v_h$
at $\gamma=1$. The full circles represent the extrapolated $N\to\infty$ values and
the dashed line is the best fit of those points.
Right panel: the same as in the left panel, but with $\gamma=0.1$.}
\end{figure}

In order to verify the ZDZ scaling law of the excitation density, 
we plot $\nu$ as function of the sweep
velocity $v_h$ in Fig.\ref{f.nu_vs_vh}. 
The data extrapolated to the thermodynamic limit
agrees well with the expected scaling law $\nu\propto \sqrt{v_h}$ 
independently of $\gamma$. 
In particular for the Ising model we find $\nu=0.11(4) \sqrt{v_h}$, while previous analytical studies based of a LZ~\cite{Dziarmaga05} and perturbative~\cite{Polkovnikov05} approximated approaches overestimated it.
With decreasing $\gamma$, the minimum gap decreases linearly in 
$\gamma$ 
while the density of state increases as $1/\gamma$. 
The combined effect results in
an overall enhancement of the number of excitations in the system and in the thermodynamic limit the density of excitations
increases as $\sim 1/\gamma$. This is well verified by comparison of
the curves in
Fig.\ref{f.nu_vs_vh} for $\gamma=1$ (left panel) and
$\gamma=0.1$ (right panel).

\begin{figure}
 \centering
\includegraphics[width=5cm]{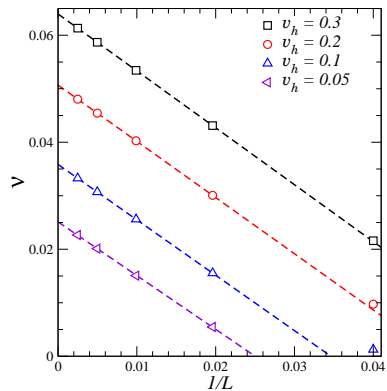}
 \caption{\label{f.extrap}
Extrapolation of the density of excitations to the thermodynamic
limit for $\gamma=1$. Symbols: $\nu(v_h,L)$; dashed lines: regression lines.}
\end{figure}

\begin{figure}
 \centering
\includegraphics[width=7cm]{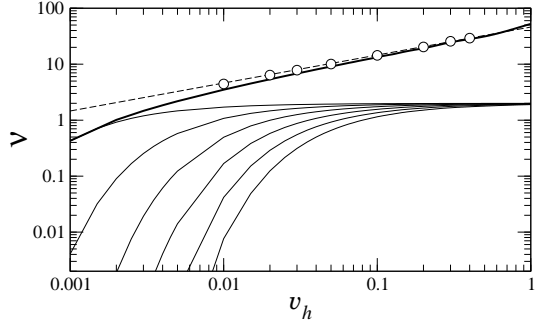}
 \caption{\label{f.single_mode}
Single mode contribution to the excitation density for $L=401$ 
and $\gamma=1$. Bold line:
total excitation density; thin lines: contribution of the first 6 modes from
the lowest (in energy) to the highest one (from top to bottom).}
\end{figure}

One can wonder whether the ZDZ scaling $\nu\propto \sqrt{v_h}$ is a
pure many-body effect or determined already by the first excited
state alone. In order to answer this question it is important first to
observe in Fig.\ref{f.extrap} the extrapolation of $\nu$ for $1/L \to
0$. The thermodynamic limit is reached with a perfect linear behavior
in $1/L$ as expected by the constant value $\frac{1}{2\gamma}$ of 
the DOS at low energy\footnote{The DOS in the low energy limit at $h=1$ reads 
$\delta_{\epsilon\to 0}\sim\frac{1}{2\gamma}
 + \frac{4\gamma^2-3}{64\gamma^5}\epsilon^2 ~.$
}. 
However, this scaling holds only for large enough values of $v_h L$. 
To be precise, a sufficiently large
number of excited states must be taken into account
in order to get the correct thermodynamic limit of $\nu$.
Even though the first excited state certainly gives the main
contribution to the overall excitation density at low sweep velocities,
Fig.\ref{f.single_mode} shows that 
all the levels between the lowest excited state up to the sweeping energy
$\epsilon_v=v_h$ are necessary for the ZDZ scaling to show up.
At fixed $v_h$ and in the limit $L\to \infty$ the number of levels 
in this interval is $\sim\frac{1}{2\gamma}(v_h - 2\pi \gamma/L)$.
So, we conclude the effect to be essentially due to many-body physics 
rather than to a simple level crossing.

The last important issue of this section is to examine how the
mesoscopic (finite) dimension of the spin chain affects the ZDZ
scaling behavior and eventually the adiabatic passage through a quantum
phase transition. In fact, two competing effects arise when chains
of finite length $L$ are considered:
\begin{itemize} 
\item on the one hand, finite $L$ means
finite gap $\Delta_L$ at $h=h_{\rm c}$ and the fidelity of the
adiabatic passage is bounded by the ZDZ scaling law;
\item on the other hand, finite $L$ also implies large 
mesoscopic fluctuations of the excitation number produced during the sweep, 
i.e. the average quantities can lose their importance in favor of 
higher moments. 
\end{itemize}
To address this question,
we evaluated the mean square fluctuations $\delta_\nu^2$
of the average density of excitations $\nu$
\begin{eqnarray}\label{flucts}
\delta^2_\nu &\equiv& \frac{1}{L^2} \langle GS | U^\dagger(t) 
\sum_{kk'} \eta_k^\dagger \eta_k \eta_{k'}^\dagger \eta_{k'} \,
U(t) | GS \rangle -\nu^2\nonumber\\
    &=& \frac{1}{L^2}\sum_k \nu_k (1-\nu_k)~.
\end{eqnarray}
\begin{figure}
    \begin{center}
    \includegraphics[width=7cm]{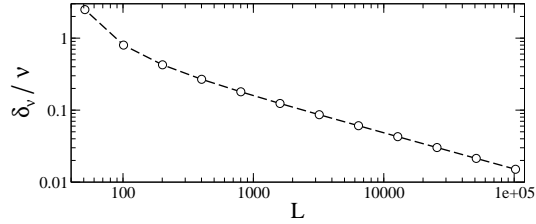}
    \end{center}
\caption{
Relative fluctuations of the excitation density versus $L$ in the
Ising model $\gamma=1$ and sweeping velocity $v_h=0.05$. The largest
lattice size is $L=102401$.\label{f.Er}}
\end{figure}
The relative fluctuations $\delta_\nu/\nu$ are plotted in
Fig.~\ref{f.Er} for the quantum Ising model, i.e. $\gamma=1$, and
a moderate sweeping velocity. The striking feature that emerges from
this analysis is the high sensitivity of the relative fluctuation to
the size of the system. Even for rather large lattice size ($L=401$)
the relative weight of $\delta_\nu$ is as large as 18\% and rapidly
increase up to 80\% for $L=101$. When the rate of the passage is
given, small lattice size implies few excitations on average, but also
very large fluctuations in their number and thus a lack of information
about the final state.

Let us remark that the ZDZ scaling emerges correctly only in the
thermodynamic limit where the relative fluctuations are strongly
suppressed. Thus, even if the dynamics of the
model~(\ref{Hamiltonian}) is governed by
the Landau-Zener tunneling of non-interacting particles, the ZDZ scaling
is the proper signature of the quantum critical point crossed during
the sweep rather than the result of single particle tunneling
processes.

\section{Robustness of the adiabatic passage in the presence of noise}
\label{noise}

A real experimental setup is subject to noise, another source that may 
corrupt the adiabatic passage fidelity. 
Here, we examine the case of a noisy driving field,
$h_\xi(t) = h(t) + \xi(t)/2$, where $\xi(t)$ is a stochastic field.
The simplest choice of a white spectrum for $\xi(t)$
(i.e.  $\expect{\xi(t)}=0$ and $\langle \xi(t)\xi(0)\rangle = \Gamma \delta(t)$) 
will be
enough for understanding the main features due to noise and decoherence. 
The effect of the noise is that a diagonal
stochastic term enters the Hamiltonian (\ref{H^e_pauli}), leading to
\begin{equation}
\label{e.Hkdiss} H^\xi_k=-\cos{\phi_k} \id_2 +
\Big[a_k+\frac{\xi(t)}{2}\Big] \sigma^z + b_k \sigma^y ~.
\end{equation}
In order to study the dynamics of such a system one has to deal with
the density matrix formalism and to solve the Lindblad
equation~\cite{lindblad}
\begin{equation}
\label{e.lindblad} \dot\rho_k(t) = -i
[H^{even}_k,\rho_k(t)]-\Gamma[\sigma^z,[\sigma^z,\rho_k(t)]]~,
\end{equation}
that can be rewritten as
\begin{equation}
\label{e.diffLk}
\dot{\vec{\rho}}_k = {\cal L}_k \vec{\rho}_k~,
\end{equation}
where $\vec{\rho}_k$ is the Bloch vector defined by $\rho_k =
\frac12 (\id + \vec{\rho}_k\cdot \vec{\sigma})$. The Lindbladian
here takes the form
\begin{equation}
{\cal L}_k =  \Matrix{ccc}{-\Gamma & -2 a_k & 2 b_k \\
2 a_k & - \Gamma & 0 \\
-2 b_k & 0 & 0}~.
\end{equation}

\begin{figure}
 \centering
\includegraphics[width=70mm]{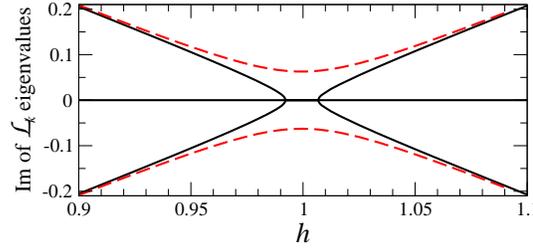}
 \caption{\label{f.Laut}
Imaginary part of ${\cal L}_k$ eigenvalues vs. $h$
for $\gamma{=}1$, $k{=}0.99 \pi$, and  $\Gamma{=}0.15$. The dashed lines represent the noiseless spectrum for the same values
of the parameters.}
\end{figure}

The stochastic noise induces a nonhermitian perturbation to
the original Hamiltonian~\eref{Hamiltonian}. The presence of such
kind of perturbations can radically modify the physics of the
unperturbed (hermitian) system, in particular when the latter has 
degeneracy or quasi-degeneracy points, as is the case here: 
degeneracy or quasi-degeneracy points typically evolve into branch-points
of nonhermitian degeneracies~\cite{berry}. 
Indeed, two eigenvalues (and the two eigenvectors) of
the fermionic system~\eref{e.Hkdiss} merge in a finite interval of $h$,
as compared with a single point without noise.
This is seen in Fig.~\ref{f.Laut} where the imaginary part of
eigenvalues of the Lindbladian ${\cal L}_k$ versus the field $h$ are
plotted.

Analysis of the noisy case admits to answer two fundamental questions:
\begin{itemize}
\item {\em Is it still
possible to discriminate the adiabatic from a sudden regime in the presence 
of noise?}
\item {\em How robust is the KZ mechanism with respect to noise?}
\end{itemize}

\begin{figure}
 \centering
\includegraphics[width=65mm]{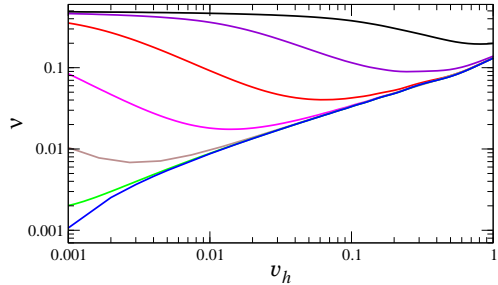}
 \caption{\label{f.etak_diss} Kink density vs. $v_h$ for $\gamma{=}1$, $L{=}401$, 
and different values of the dissipative coupling: From bottom to top,
 $\Gamma=0,10^{-6},10^{-5},10^{-4},10^{-3},10^{-2},10^{-1}$.}
\end{figure}

Solving (numerically) Eq.~(\ref{e.diffLk}) we evaluate
the kink density (excitation density) of the system using
Eq.~(\ref{e.nu_k}). It is shown in Fig.~\ref{f.etak_diss} 
for the Ising model ($\gamma=1$) after a sweep of the field from
$h_0=10$ to $h_f=0$.
The excitation density of the system increases considerably with the noise
strength $\Gamma$ up to the saturation value $1/2$ (the maximum
number of kinks is $L/2$). 
Besides this, there are two new and striking features:
\begin{enumerate}
\item the considerable noise instability of the adiabatic approximation 
even if $\Gamma$ is orders of magnitude smaller than the excitation gap; 
\item the non monotonic behavior of the excitation density with the quench rate $v_h$.
\end{enumerate}
In order to understand this somehow counterintuitive behavior,
all relevant time scales for the adiabatic passage need to be taken 
into account: the relaxation time $\tau=1/\Delta$, which is the only relevant
scale in the noiseless case, and the decoherence time $\tau_\varphi$ [in
the white noise case $\tau_\varphi=1/(2\Gamma)$]. Both time
scales have to be compared with the sweeping time $T=(h_0-h_f)/v_h$.
In order to faithfully drag a quantum state by the drive, 
the sweeping rate has to be much slower than the gap of the
system, $v_h\ll\Delta$, but, on the other hand, it has to be much
faster than the dephasing rate, $v_h\gg\Gamma$. Then, the condition
for the success of the adiabatic passage is that $\Gamma\ll v_h\ll
\Delta$.

\begin{figure}
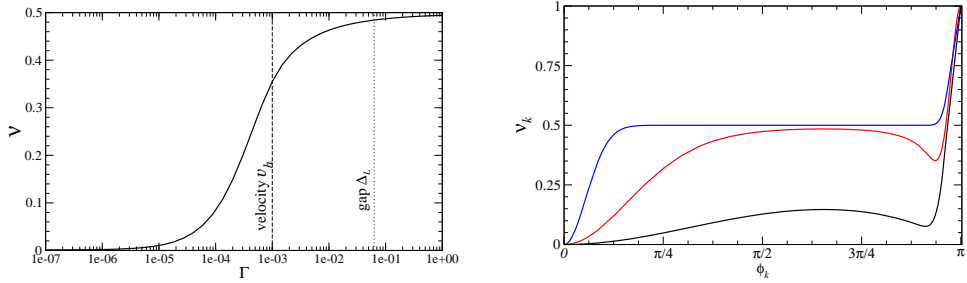

 \centering
 \subfigure
 {\includegraphics[width=60mm]{fig10left.eps}}
 \hspace{5mm}
 \subfigure
{\includegraphics[width=60mm]{fig10right.eps}}
\caption{\label{f.etak_k} Left panel: Kink density vs. $\Gamma$ for
$v_h=0.001$, $\gamma=1$ and $L=401$. 
Right panel: Single mode excitation density vs. $\phi_k$ for different values
of the ratio $\Gamma/v_h=0.1,1,10$ (from bottom to top,
respectively).}
\end{figure}

This is nicely seen from the excitation probability as a
function of $\Gamma$ (Fig.~\ref{f.etak_k}, left panel) and explains
the observed non-monotonicity. The same feature is resolved in terms 
of modes $\phi_k$ in the right panel of the same figure. 
When $\Gamma$ is of the order of $v_h$ or even larger,
most of the modes lose their quantum coherence during the
sweep irrespective of the velocity. Eventually, the excitation 
density of each mode saturates to the value 1/2. The relevance of the 
appearing of the decoherence time compromises the distinction between 
the adiabatic and the sudden region.
\begin{figure}
 \centering
\includegraphics[width=60mm]{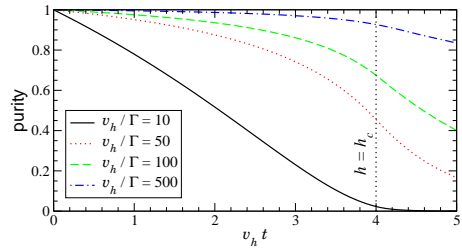}
\caption{\label{f.purity} Purity of the state, $\Tr\rho^2$, during the 
sweep from $h(0)=5$ to $h(T)=0$ for different values of $v_h$ and $\Gamma$. 
The curves depend only on the ratio $v_h/\Gamma$.}
\end{figure}
The noise progressively degrades the state of the system into a mixed one. 
This shows up in Fig.\ref{f.purity} looking at the behavior of the purity 
defined as $\Tr\rho^2$, that we find to depend solely on the 
velocity-to-noise ratio $v_h/\Gamma$. The rate of degradation of the 
state during the sweep is constant only far from the critical point 
and this is particularly evident in the case of low velocity-to-noise 
ratio (full line) where the state mixes up almost completely before reaching 
$h(t_{\rm c})=h_{\rm c}$. For large enough values of $v_h/\Gamma$, 
approaching the critical point the  mixing 
rate of the state increase as shown by the flex-point 
in the purity close to $h_{\rm c}$. 
This puts clear emphasis on the various facets of noise: besides
the mere relaxation-excitation process the main effect consists in
decoherence, i.e. mixing of the density matrix.
The latter effect has been discarded in a previous work on this
subject~\cite{Childs}. We conclude that the (quasi-)degeneracy point 
not only gives a limit to the rate the parameters of the Hamiltonian 
may change, but more importantly, it amplifies the effect of decoherence.

\section{Conclusions}\label{concl}

We have analyzed the dynamical crossing of a quantum phase transition
analytically for an ideal clean system and numerically in the presence
of white noise. Our focus was to determine the robustness of the 
adiabatic passage trough a quasi degeneracy point of a many-body system 
when one considers the effects of finite sweeping period, 
mesoscopic fluctuations and stochastic forces.

We found a surprisingly high predictive power of the Landau-Zener formula
for the final excitation density, when several modes are taken into account.
Its validity exceeds significantly the validity range attributed to it
in Ref.~\cite{Dziarmaga05} when the number of Landau-Zener modes to be taken
into consideration are determined by the density of states and
the characteristic energy scale of the sweep. This number
diverges in the thermodynamic limit and it is only in this limit
that the ZDZ-scaling~\cite{ZurekDornerZoller} manifests.
In this sense is this scaling essentially due to collective many-particle
physics rather than to the Landau-Zener tunneling of few two-level
systems. In addition huge fluctuations around the average
excitation density dominate the state during the sweep, in particular for 
not too large systems. 
This tells that the physics of the transition
is not entirely captured by the excitation average and its square root 
ZDZ-scaling. 
The implication of this fact deserve further analysis in order 
to understand its implications for AQC. 
In fact, large fluctuations on the average number of excitations implies 
a loss of information on the final state of an AQC process.

The presence of noise corrupts the success of a possible adiabatic passage
for low sweeping velocities due to the opening of a non-hermitian degeneracy
leading to a ``critical region'' rather than just a singular point.
The same phenomenon appears also in presence of just an avoided level crossing
or a quasi-degeneracy of the spectrum.
This leads to an all-over increase of the average excitation density, but it is shown that the most devastating effect of the noise
on the transfer fidelity across a non-hermitian degeneracy region is the
decay of the state purity that is emphasized by the passage through the degeneracy.
Therefore, one has to consider a second time scale besides $1/\Delta$: 
the dephasing time. These two time scales are the upper and lower 
bound for the sweeping rate. Thus there exists an optimal velocity for 
each noise level that minimizes the fidelity loss.

Finally let us comment on different kind of noise couplings that could 
be of interest for future studies. In fact, one can wonder how the present 
conclusions may change if the stochastic variables couple with 
$S^{x(y)}$, breaking the symmetry of the system. Even though the 
noise coupling we considered preserves the partity symmetry of the Hamiltonian 
(but not the time reversal one), it causes both relaxation 
(for $h\neq h_c$) and decoherence. In this sense, it provides
a complete phenomenology. Moreover, the main feature we find, namely 
the enhancement of the mixing up rate of the state, should be independent 
on the kind of noise coupling. With this respect, it would be interesting
to study colored noise that can strongly modify the dephasing time;
it should also have an effect on the white-noise induced 
increase of the excitation density with time during all the sweep.

\ack

The authors would like to acknowledge L. Amico, E. Paladino, A. Polkovnikov 
and J.~Siewert
for useful discussions and R. Fazio for discussions and hospitality
at the SNS Pisa. A.F. acknowledges the PRIN2005029421 project for financial support 

\appendix
\section{Exact diagonalization of the XY model in transverse field}

In order to diagonalize the one-dimensional XY model in transverse
magnetic field (\ref{Hamiltonian}) we follow
Refs.~\cite{McCoy1,LiebSchulzMattis,Pfeuty} and perform the
substitution
\begin{eqnarray}\label{hard-core-bosons}
a^\dagger_j&:=&S^x_j + i S^y_j\\
n_j&:=&a^\dagger_ja^{}_j=\frac{1}{2}+S^z_j.
\end{eqnarray}
The resulting (hard-core bosonic) Hamiltonian becomes
\begin{equation}
\label{Hamiltonian:hard-core-bosons}
\fl
H=-\frac{1}{2}\left[\sum_{j=1}^{L=2 N+1}
    (a^\dagger_j a^{}_{j+1}+ {\rm h.c.})
    + \gamma (a^\dagger_j a^\dagger_{j+1} + {\rm h.c.})\right]
    - h(t) \sum_{j=1}^{L=2 N+1} n_j + \frac12 Lh(t)\,.
\end{equation}
Here we restricted ourselves to an odd number of sites $L=2N+1$ 
with integer $N$. Though this is not crucial for the exact diagonalization 
of the Hamiltonian, this becomes equivalent to the restriction
to an odd total number of particles -- and hence no boundary phase -- 
when starting from a fully polarized state at very large magnetic field.

The Jordan-Wigner transformation
\begin{eqnarray}
\label{JW}
c^\dagger_j&:=&a^\dagger_j \exp\left(i \pi \sum_{l=1}^{j-1} n_j\right)\\
c^\dagger_j c^{}_j &=& a^\dagger_j a^{}_j = n_j
\end{eqnarray}
further transmutes the Hamiltonian into fermionic form
\begin{equation}\label{Hamiltonian:fermions}
\fl
H=-\frac{1}{2}\left[\sum_{j=1}^{L=2 N+1}
    (c^\dagger_j c^{}_{j+1}+ {\rm h.c.})
    + \gamma (c^\dagger_j c^\dagger_{j+1} + {\rm h.c.})\right]
    - h(t) \sum_{j=1}^{L=2 N+1} n_j + \frac12 Lh(t)
\end{equation}
Subsequent Fourier transformation
\begin{equation}\label{FT}
f^\dagger_k:=\frac{1}{\sqrt{L}}\sum_l \exp(-i k l)\, c^\dagger_l
\end{equation}
completely decouples the Hamiltonian, so that $H=\oplus_{|k|} H_k$,
where $H_k$ acts in the 4-dimensional Hilbert space
$\{\ket{0}_{k,-k},\ket{k,-k};\ket{k},\ket{-k}\}$. Due to the parity
symmetry, it further decouples into $H_k=H_k^{odd}\oplus
H_k^{even}$. Where $H_k^{odd}=-\cos{\phi_k}\id_2 $, while in the
even-occupation Hilbert space and basis $\{\ket{0},\ket{k,-k}\}$ one
obtains
\begin{eqnarray}
\label{H-even}
H_k^{even}&=& \Matrix{cc}{
h(t) & - i\gamma\sin{\phi_k}\\
i\gamma\sin{\phi_k} & -(2\cos{\phi_k}+h(t))} \nonumber\\
&=& - \cos{\phi_k}\,\id_2 + \Matrix{cc}{
\cos{\phi_k}+h(t) & -i\gamma\sin{\phi_k}\\
i\gamma\sin{\phi_k} & -(\cos{\phi_k}+h(t)) }\equiv \nonumber\\
&\equiv& - \cos{\phi_k}\,\id_2 + \Matrix{cc}{
a_k  & -i b_k \\
i b_k & -a_k }\; .
\end{eqnarray}
Expressed in Pauli matrices this is
$$
H_k^{even}=-\cos{\phi_k} \id_2 + a_k \sigma^z + b_k \sigma^y
$$
The eigenvalues and the ground state of the nontrivial second part
are given by Eqs.(\ref{Ek}) and (\ref{e.GS}), respectively.

\section{Exact time evolution of the XY model in transverse field}

The Heisenberg equation (\ref{e.HE}) for the time evolution operator
has been solved in Ref.~\cite{McCoy1} to obtain the exact time
evolution of correlation functions of the model. Using the notation
$$
U^{(k)}=\Matrix{cc}{
U^{(k)}_{11} & U^{(k)}_{12}\\
U^{(k)}_{21} & U^{(k)}_{22}}
$$
we obtain the following set of differential equations
\begin{eqnarray}\label{U-evolution}
i\ddot{U}^{(k)}_{11}&=&\left(\dot{a}_k-i \Lambda_k^2\right) U^{(k)}_{11} \\
i\ddot{U}^{(k)}_{22}&=&\left(-\dot{a}_k-i
\Lambda_k^2\right) U^{(k)}_{22}
\end{eqnarray}
>From these equations and the initial conditions one concludes
$U_{22}^{(k)}= U_{11}^{(k)*}$ and $U^{(k)}_{21}=-U^{(k)*}_{12}$ and
the missing amplitudes are extracted as
\begin{eqnarray}
U^{(k)}_{12}&=& \frac{1}{b_k}\left(\dot{U}^{(k)}_{22}
            -ia_k {U}^{(k)}_{22}\right)\\
U^{(k)}_{21}&=& \frac{1}{b_k}\left(-\dot{U}^{(k)}_{11}
            -ia_k {U}^{(k)}_{11}\right)\quad .
\end{eqnarray}
The differential equation (\ref{U-evolution}) is exactly solvable if
$a_k$ is polynomial in $t$~\cite{VOROS} and even for particular
cases of transcendental dependence of $t$~\cite{McCoy1}. For the
simplest case $h(t)=h_0 - v_h t$, with $v_h=(h_0-h_1)/T$, the solution is
\begin{eqnarray*}
\fl
        U_{11}^{(k)}=&{\rm e}^{-\frac{i}{2}v_ht^2-
        (h_0+\cos{\phi_k}) t}\left[ C_1
        {\rm H}[i\gamma^2\frac{\cos{2\phi_{k}}-1}{4v_h},
        -\frac{1+i}{\sqrt{2v_h}}(h(t)+\cos{\phi_k})]\right.\\
\fl
        &\left. \quad +C_2 {\rm
        F}[i\gamma^2\frac{\cos{2\phi_k}-1}{8v_h},\frac{1}{2},
        i 2v_h(h(t)+\cos{\phi_k}))^2]\right] \nonumber\\
\fl
        U_{22}^{(k)}=&{\rm e}^{-\frac{i}{2}v_ht^2-
        (h_0+\cos{\phi_k}) t}\left[ C_1 {\rm
        H}[i\gamma^2\frac{\cos{2\phi_k}-1}{4v_h}-1,
        -\frac{1+i}{\sqrt{2v_h}}(h(t)+\cos{\phi_k})]\right.\\
\fl
        &\left.\quad +C_2 {\rm
        F}[i\gamma^2\frac{\cos{2\phi_k}-1}{8v_h}-\frac{1}{2},\frac{1}{2},
        i 2v_h(h(t)+\cos{\phi_k})^2] \right]\nonumber
\end{eqnarray*}
In our notation $F[x,y,z]$ is the confluent hypergeometric function
and has the series expansion
$$
F[x,y,z]=\sum_{k=0}^\infty \frac{(x)_k}{(y)_k} \frac{z^k}{k!}
$$
with $(a)_k:=a (a+1) \cdots (a+k)$ the Pochhammer symbol. Its pole
structure is that of the gamma function $\Gamma(y)$ such that
$F[x,y,z]=\Gamma[y] \tilde{F}[x,y,z]$ with the entire function
$\tilde{F}[x,y,z]$. The generalized Hermite polynomials
$H_\nu[z]=:H[\nu,z]$ are expressed in terms of the confluent
hypergeometric function
$$
H[\nu,z]=2^\nu\sqrt{\pi}\left[
\frac{F[-\frac{\nu}{2},\frac{1}{2},z^2]}{\Gamma[\frac{1-\nu}{2}]}
    -\frac{F[\frac{1-\nu}{2},\frac{3}{2},z^2]}{
        \Gamma[\frac{-\nu}{2}]}\right]\,.
$$
At integer values of $\nu$, this function coincides with the Hermite
polynomials.

\end{document}